\documentclass[letterpaper,english,twocolumn,aps,prl,amsfonts,amssymb,amsmath,longbibliography,superscriptaddress]{revtex4-2}
\usepackage[utf8]{inputenc}
\usepackage{graphicx}
\usepackage{bbold}
\usepackage{xcolor}
\usepackage[normalem]{ulem}

\newcommand{\be}{\begin{equation}}
\newcommand{\ee}{\end{equation}}

\newcommand{\bspl}{\begin{split}}
\newcommand{\espl}{\end{split}}
\newcommand{\bea}{\begin{eqnarray}}
\newcommand{\eea}{\end{eqnarray}}

\def\a{\alpha}
\def\b{\beta}

\def\o{\omega}

\def\O{\Omega}

\def\ra{\rightarrow}
\def\up{\uparrow}

\def\down{\downarrow}

\def\pd{\partial}

\def\bj{{\bf j}}

\def\bA{{\bf A}}

\def\nn{\nonumber}
\def\lb{\label}
\def\pref#1{(\ref{#1})}

\newcount\bozza \bozza=0
\ifnum\bozza=1
\newdimen\shift \shift=-2truecm
\def\lb#1{%
{\label{#1}\rlap{\kern\shift{$\scriptstyle#1$}}}}
\else\def\lb#1{\label{#1}} \fi

\begin{document}
	
	\title{Coherent driving of displacive Higgs fluctuations in  superconductors}

\author{Jacopo Fiore}
	\email{jacopo.fiore@uniroma1.it}
	\affiliation{Department of Physics, ``Sapienza'' University of Rome, P.le
		A. Moro 5, 00185 Rome, Italy}
    \affiliation{Institute for Theory of Statistical Physics, RWTH Aachen University, Aachen, Germany}
	\author{Irene Zanotti}
	\affiliation{Department of Physics, ``Sapienza'' University of Rome, P.le
		A. Moro 5, 00185 Rome, Italy}
        \author{Kota~Katsumi}
\affiliation{William H. Miller III , Department of Physics and Astronomy, The Johns Hopkins University, Baltimore, MD, 21218, USA} 
\affiliation{Center for Quantum Phenomena, Department of Physics, New York University, New York, New York 10003, USA}
\author{N.~P.~Armitage}
\affiliation{William H. Miller III , Department of Physics and Astronomy, The Johns Hopkins University, Baltimore, MD, 21218, USA} 
	\author{Claudio Castellani}
	\affiliation{Department of Physics, ``Sapienza'' University of Rome, P.le
		A. Moro 5, 00185 Rome, Italy}
    \affiliation{ISC-CNR, ``Sapienza'' University of Rome, P.le
		A. Moro 5, 00185 Rome, Italy}
\author{Goetz Seibold}
\affiliation{Institut f\"{u}r Physik, BTU Cottbus-Senftenberg, P. O. Box 101344, 03013 Cottbus, Germany}
	\author{Lara Benfatto}
	\email{lara.benfatto@roma1.infn.it}
	\affiliation{Department of Physics, ``Sapienza'' University of Rome, P.le
		A. Moro 5, 00185 Rome, Italy}

   \begin{abstract}
    Intense phase-stable terahertz (THz) laser pulses can  drive collective modes coherently via multi-photon excitation pathways in a manner different than the standard resonant mechanism operative in linear response. Here we show that in superconductors the nonlinear optical response can be enhanced when excited quasiparticles activate a (non-resonant) static displacement of the superconducting order parameter, in full analogy with the displacive excitation of coherent phonons in opaque materials. 
    By combining numerical simulations with analytical results we demonstrate that the displacive mechanism to excite the Higgs mode is operative in both $s$-wave and $d$-wave superconductors. We validate this prediction experimentally by the temperature dependence of the phase of the nonlinear first harmonic in superconducting $s$-wave NbN. We also discuss how the order-parameter relaxation at large times, which can be experimentally accessed via pump-probe protocols is connected to energy-dissipative processes. Our results offer a novel perspective on the ability of intense THz fields to measure, and eventually control, the parametric dependence of the optical response on collective degrees of freedom.
   \end{abstract}

	\date\today
	\maketitle

	{\bf Introduction}
Light-based spectroscopies with weak fields provide an ideal experimental realization of the fluctuation-dissipation theorem. In this linear regime, the response is maximized at resonance, i.e.  when the excitation frequency is tuned to the characteristic fluctuation energy of a selected degree of freedom~\cite{fabrizio_book}. However, when the intensity of the light pulses is sufficient to activate nonlinear optical processes, the standard paradigm of resonantly driving a collective excitation coexists with phenomena that seem to defy a quasi-equilibrium description, such as the emergence of long-lasting signals in pump-probe schemes~\cite{giannetti_Adv.Phys.16, yang_NatRevMater23}.
  Here, we demonstrate that both types of phenomena are intrinsically encoded within the nonlinear optical response computed at equilibrium. By benchmarking this mechanism in disordered superconductors, we show that the enhanced nonlinear response is tied to the off-resonance driving of {\em static} fluctuations of the superconducting (SC) order parameter, commonly known as the Higgs mode ~\cite{shimano_Annu.Rev.Condens.MatterPhys.20}. By combining numerical simulations of a prototype model for $s$-wave superconductors with analytical calculations, we prove that the nonlinear current $j^{NL}$ driven by a nearly monochromatic pulse with a central frequency $\Omega$ is dominated by the term:
\be
\lb{pnlsc}
j^{NL}(\omega+\Omega)\sim  \frac{\pd \chi(\O)}{\pd \Delta} \delta \Delta(\o) A(\O),
\ee
where $A(\Omega)$ is the electromagnetic vector potential and $\chi(\Omega)$ is the linear electronic susceptibility, which depends {\em parametrically} on the SC order parameter $\Delta$. The amplitude fluctuations $\delta\Delta(\omega)$, in turn, scale as:
\be
\lb{eliash}
\delta \Delta(\o)\sim \frac{2i\text{Im}\chi(\Omega)}{\omega+i\eta}H(\omega) A(\O)A(\o-\O),
\ee
where $\eta$ generically denotes the inelastic quasiparticle scattering rate, $H(\omega)$ is the Higgs-mode propagator and $\o\sim 0$ for nearly monochromatic pulses. 
For a dirty $s$-wave superconductor, the linear susceptibility $\chi(\Omega)$ is fully gapped at low $T$ for $\Omega<2\Delta$~\cite{mattis_Phys.Rev.58,zimmermann_PhysicaC:Superconductivity91}. The onset of absorption at $\Omega>2\Delta$ is responsible for a strong enhancement of $j^{NL}$. 
This enhancement is driven both by the sharp increase in $\partial \chi / \partial \Delta$ at the optical gap and by the activation of the amplitude fluctuations $\delta\Delta$ in Eq.\ \pref{eliash}, whose static component diverges at $\eta=0$. The emergence of this large ``rectified'' component is physically analogous to the displacive excitation of coherent phonons in insulators~\cite{merlin_ssc97,dresselhasu_prb92,merlin_prb02}, which is observed  when the frequency of the light pulse exceeds the band gap~\cite{dresselhaus_apl90,merlin_prb02,bossini_prb25}. In other words, Eq.\ \pref{eliash} provides a quantitative description of  the standard physical picture of a quench, where the injection of quasiparticles 
across a gap alters the free-energy landscape, shifting the equilibrium atomic coordinates in the case of phonons, or the static order parameter in the case of superconductors. Near the critical temperature $T_c$, this process can even lead to the enhancement of the SC order parameter under continuous microwave irradiation, a phenomenon widely known as the Eliashberg effect~\cite{eliashberg_JETPLett.70,derendorf_Phys.Rev.B24}. Numerical simulations for a $d$-wave order parameter confirm the $1/\eta$ scaling of the static fluctuation $\delta\Delta(0)$ predicted by
Eq.\ \pref{eliash}. However, in the $d$-wave case, low-energy excitations below twice the gap maximum smooth out the derivative $\partial \chi / \partial \Delta$, thereby suppressing the Higgs-mode contribution to the nonlinear current $j^{NL}$ described by Eq.\ \pref{pnlsc}. 

The mechanisms described by Eqs.\ \pref{pnlsc} and \pref{eliash} yield direct and observable experimental signatures. The first is the thermal enhancement of the nonlinear first-harmonic (FH) current $j^{NL}(\Omega)$ at the temperature $T^*$ where $2\Delta(T^*) = \Omega$, as recently measured by means of THz two-dimensional coherent spectroscopy (2DCS)~\cite{mukamel_95,hamm_11,cundiff_PhysicsToday13,smallwood_LaserPhotonicsRev.18} in $s$-wave NbN ~\cite{katsumi_Phys.Rev.Lett.24}. In this work, we present additional measurements of the FH phase as a function of temperature, demonstrating that its shift at $T^*$ directly tracks the phase shift of $\partial \chi / \partial \Delta$ in Eq.\ \pref{pnlsc}. 
A second experimental fingerprint of this displacive Higgs excitation manifests in time-resolved protocols, since the $\omega \sim 0$ enhancement of the Higgs modulation in Eq.\ \pref{eliash} generates a rectified pump-probe signal that decays on a timescale of $1/\eta$, reminiscent of the transient responses observed in superconductors under intense THz driving\cite{demsar_prl13,demsar_prl23,matsunaga_Phys.Rev.Lett.13,matsunaga_Science14,chaudhuri_25}.  We show that the rate $\eta$ in Eq.\ \pref{eliash} directly measures the energy-dissipation rate, in contrast to the momentum-dissipation rate probed by linear response~\cite{barbalas_Phys.Rev.Lett.25,chaudhuri_25}. This result is established by demonstrating a formal analogy between Eq.\ \pref{eliash} and the standard description of power absorption under an external electromagnetic perturbation~\cite{fabrizio_book}.
The key requirement is that the collective mode driven by the difference-frequency mixing of two pulses at $\Omega$ and $-\Omega + \omega$ behaves as a conserved quantity in the quasiparticle basis, a condition that is always satisfied by the amplitude mode in a superconductor. More generally, our results offer a novel perspective on how nonlinear optical spectroscopy can access and exploit excitation and relaxation pathways that linear probes are blind to.

{\bf Nonlinear first harmonic}.  
We are interested in determining the nonlinear current $\bj^{NL}$ in a superconductor in response to a time-dependent vector potential $\bA$. Firstly, focusing on experiments with multicycle monochromatic pulses~\cite{matsunaga_Science14,matsunaga_Phys.Rev.B17,kovalev_Phys.Rev.B21,chu_NatCommun20,chu_NatCommun23,yuan_Sci.Adv.24,grasset_npjQuantumMater.22,kim_Sci.Adv.24,katsumi_Phys.Rev.Lett.24,katsumi.Phys.Rev.Lett.25} we represent the external gauge potential as ${\bf A}(t)=\bA_0\cos(\Omega t)$. In a centrosymmetric system the leading nonlinear current scales with the third power of the gauge field, $j^{NL}\sim \chi^{(3)}A^3$, which admits both $\Omega$ and $3\Omega$ harmonics thanks to the possible combinations of the spectral components at $\pm \Omega$ of the perturbation. Experimentally, $j^{NL}(3\Omega)$ can be easily isolated by measuring the third-harmonic (TH) of the electric field generated by a thin SC film~\cite{matsunaga_Science14,matsunaga_Phys.Rev.B17,kovalev_Phys.Rev.B21,chu_NatCommun20,chu_NatCommun23,yuan_Sci.Adv.24,kim_Sci.Adv.24}.  On the other hand, to isolate the FH from the much larger linear component in transmission, one can use a subtraction procedure after repeating the transmission experiment with each pulses separately, as done in Ref.\ ~\cite{katsumi_Phys.Rev.Lett.24,katsumi.Phys.Rev.Lett.25}. 

Extensive experimental work in several families of superconductors showed that in the SC state the nonlinear current is much larger than in the parent metallic state~\cite{matsunaga_Science14,matsunaga_Phys.Rev.B17,kovalev_Phys.Rev.B21,chu_NatCommun20,chu_NatCommun23,yuan_Sci.Adv.24,katsumi_Phys.Rev.Lett.18,kim_Sci.Adv.24,katsumi_Phys.Rev.Lett.24,katsumi.Phys.Rev.Lett.25}, with a thermal enhancement of $j^{NL}(3\Omega)$ when $2\Omega=2\Delta(T)$ in $s$-wave superconductors~\cite{matsunaga_Science14,kovalev_Phys.Rev.B21}. Since $2\Delta$ identifies the fluctuation frequency of both the BCS quasiparticle excitations~\cite{cea_Phys.Rev.B16,silaev_Phys.Rev.B19,seibold_Phys.Rev.B21} and the Higgs mode~\cite{tsuji_Phys.Rev.B15,schwarz_Phys.Rev.B20,tsuji_Phys.Rev.Research20,schwarz_NatCommun20,
silaev_Phys.Rev.B19,seibold_Phys.Rev.B21}, this result appears as the nonlinear counterpart of the fluctuation-dissipation theorem for linear response: the large signal at $2\Omega=2\Delta$ can be understood as a  resonant sum-frequency process, where two photons of the THz pump pulse resonantly drive a collective excitation to its characteristic frequency. On the other hand, the observation in Ref.\ ~\cite{katsumi_Phys.Rev.Lett.24} of a thermal enhancement of the FH $j^{NL}(\Omega)$ in the $s$-wave superconductor NbN at $\Omega=2\Delta(T)$ calls for a fundamentally different mechanism, which is the focus of the present manuscript. 

{\bf Model}
We focus first on a prototype SC model for the $s$-wave pairing. We consider a two-dimensional tight-binding model on a square lattice with hopping ($t_{ij}=t$) restricted to nearest neighbors $\langle ij\rangle$ only, and we include both an on-site attraction ($ -|U|$) and local random potential ($V_i$) (see e.g.\, Ref. ~\cite{ghosal01} )
\begin{eqnarray}
  \hat{H}&=&\sum_{\langle ij\rangle,\sigma}\left(t_{ij}-\mu\delta_{ij}\right) c_{i,\sigma}^\dagger c_{j,\sigma}
  -|U|\sum_{i} \hat{n}_{i,\uparrow}\hat{n}_{i,\downarrow}\nonumber\\
 &+&\sum_{i\in L,\sigma}V_i \hat{n}_{i,\sigma} \,. \label{eq:ham}
\end{eqnarray}
The inclusion of the last term, which describes static point-like impurities, is motivated by the crucial role that disorder plays in enhancing  $j^{NL}(3\Omega)$, as proven by several previous investigations~\cite{silaev_Phys.Rev.B19,tsuji_Phys.Rev.Research20,seibold_Phys.Rev.B21,benfatto_Phys.Rev.B23,haenel_Phys.Rev.B21}. In addition,  the numerical evaluation of $j^{NL}(\Omega)$ in Ref.\ ~\cite{katsumi_Phys.Rev.Lett.24} showed an even stronger impact of disorder on the FH signal. We model $V_i$ as a random variable drawn from a uniform distribution in the range $-V_0 \le V_i \le V_0$, where $i \in L$ labels a subset $L$ of lattice sites with concentration $c = |L|/N$, with $N$ the total number of lattice sites.


We compute the nonlinear current in response to ${\bf A}(t)$, coupled to the kinetic-energy term of Eq.\ (\ref{eq:ham}) via the Peierls substitution, by implementing a perturbative density-matrix approach at $T=0$. The ground-state configuration for each disorder realization is obtained by diagonalization within the Bogoliubov-de Gennes approach on finite lattices up to $24\times 24$ sites, so that we add superconductivity in a BCS-like approximation to the exact disorder eigenstates~\cite{suppl} computed for a given $V_i$ configuration. All results are then averaged over $20-30$ disorder configurations. The spectral gap $\Delta$ is  defined as the lowest positive energy eigenvalue $E_n$. With a monochromatic light pulse at $\Omega$ we can only test the $\omega=0$ limit of Eq.\ \pref{eliash}, which corresponds to evaluating the FH $j^{NL}(\Omega)$ of the nonlinear current, obtained by the combination of three photons at energies $(\Omega,-\Omega,\Omega)$. Later on we will discuss analytically the more general derivation of Eq.\ \pref{eliash} in the case of realistic light pulses with finite bandwidth. We  compute  the FH $j^{NL}(\Omega)$ by expanding the density matrix up to third order in the perturbation and providing a solution in frequency space, as a function of the exact $E_n$ eigenstates, see ~\cite{seibold_Condens.Matter23,suppl} for additional details. This approach has the twofold advantage of allowing for larger lattices than the usual time-domain solution and for a straightforward mapping of the relevant  processes into the corresponding diagrammatic structure. Our results correspond to include in the nonlinear response both the contribution of (fermionic) BCS quasiparticles and (bosonic) collective modes of the SC order parameter, computed at RPA level. To highlight the specific role of the Higgs mode, we will show results with and without the inclusion of amplitude fluctuations of the order parameter, denoted BCS+Higgs and BCS, respectively. The calculation of the exact disorder eigenstates has some advantages over a theoretical  approach aimed at including disorder in a perturbative way starting from the  momentum eigenstates. For example, a Born approximation for weak disorder requires including vertex correction to fulfill gauge invariance~\cite{silaev_Phys.Rev.B19,tsuji_Phys.Rev.Research20,tsuji_prb26}. In the exact configuration approach, the effect of elastic scattering, which governs the momentum decay, is included to all orders, and translational invariance is recovered by averaging on various configurations of disorder. The disorder-induced electronic lifetime $\tau=1/\gamma$ will be estimated from the optical conductivity in the normal state. In addition, since the mean-field Bogoliubov quasiparticles do not decay, even though elastic momentum damping is already included any additional source of damping, like the $\eta$ parameter appearing in Eq.\ \pref{eliash}, can only be provided by {\it inelastic} scattering processes, which are not explicitly computed in the present treatment.

\begin{figure}
  \includegraphics[width=\columnwidth,clip=true]{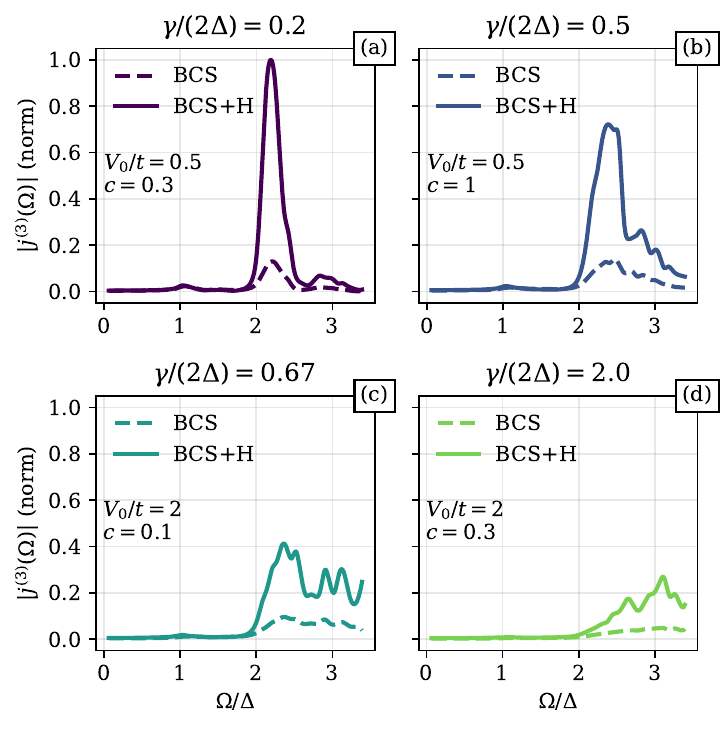}
\caption{Nonlinear first-harmonic current response $j^{(3)}(\Omega)$
    as a function of $\Omega/\Delta$, where $\Omega$ denotes the
    frequency of the applied field and $\Delta$ is the spectral
    order parameter. The different values $\gamma/(2\Delta)$ are obtained for different combinations of the impurity concentration ($c$) and impurity strength ($V_0/t$), indicated in each panel, with the same electron filling $n=0.5$. We show the nonlinear current with (solid line) and without (dashed line) Higgs fluctuations. All panels share a common scale set by the maximum value in panel (a).}  
\label{fig1}                                                   
\end{figure}  

\begin{figure}
  \includegraphics[width=\columnwidth,clip=true]{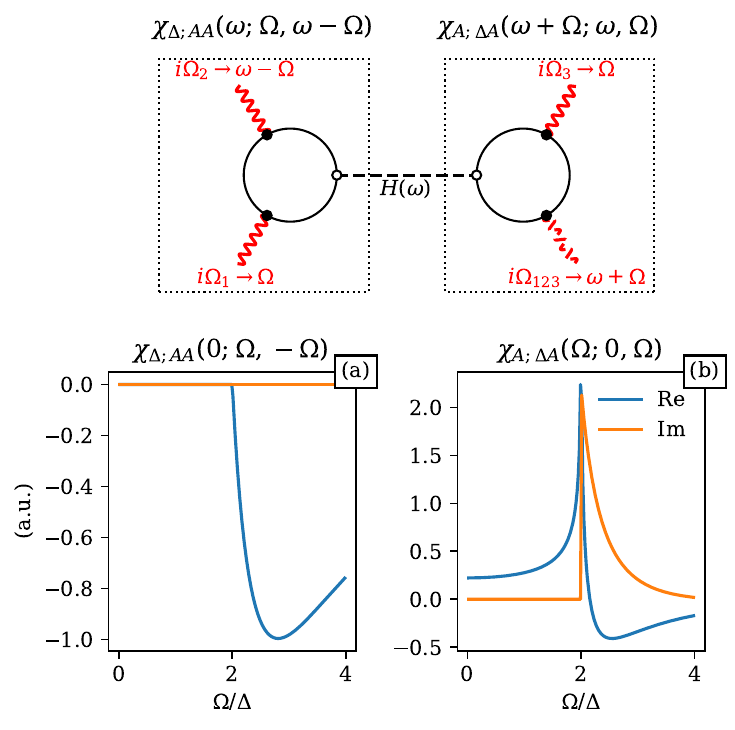}
  \caption{Top: Feynman's diagram of the amplitude-mode contribution to the nonlinear FH. Here photons are represented as wavy lines, fermions by solid lines, and  the amplitude mode by a dashed line. Filled/empty circles represent $\sigma_0/\sigma_2$ vertices in the Nambu notation (see ~\cite{suppl}). Bottom: sketch of the real and imaginary part of the left (a) and right (b) bubbles in the top panel computed at $T=0$ within the quasiparticle analytical approximation ~\cite{suppl} as a function of $\Omega$ for $\omega=0$.}     
\label{fig2}                                                   
\end{figure}  

\begin{figure}
  \includegraphics[width=\columnwidth,clip=true]{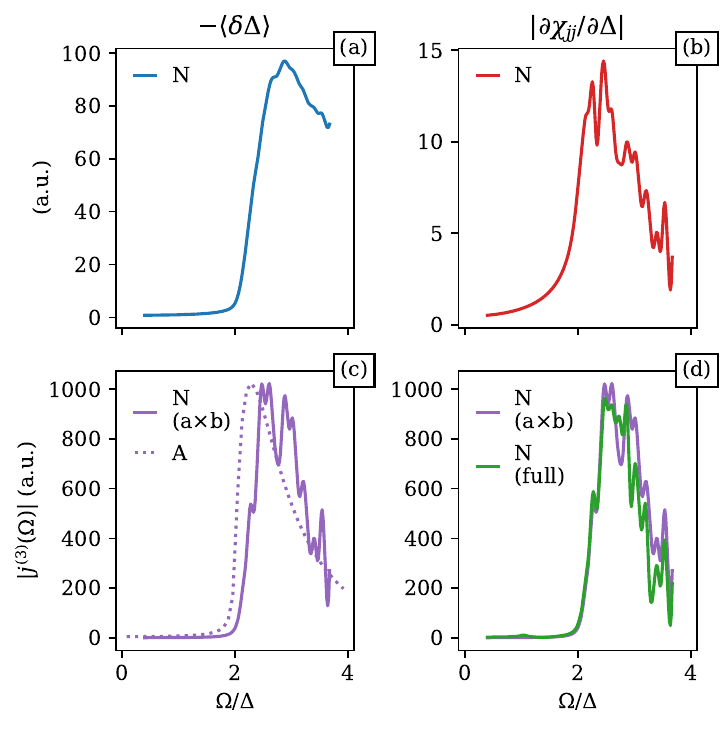}
  \caption{ (a) Static $\delta \Delta(0)$ order-parameter fluctuations and (b) derivative of the linear susceptibility $\lvert\partial \chi_{jj}/\partial\Delta\rvert$ as a function of the pump frequency $\Omega$, as obtained by numerical (N) solution of the model \pref{eq:ham}. Here we show results for impurity strength $V_0/t=0.5$ and concentration $c=1$, with electronic filling $n=0.875$.
(c) Comparisons between the product $\delta \Delta(0)(\pd\chi/\pd \Delta)$ in Eq.\ \pref{eq:japprox} obtained from the numerical results of panels (a) and (b) (solid line) and the analytical result (dotted line, denoted by A) based on the Zimmermann approximation ~\cite{suppl}, setting $\gamma/(2\Delta)\sim0.7$ and $\eta/(2\Delta)\sim0.02$ consistent with numerical simulations. Panel (d): comparison between the amplitude of the full Higgs-only FH contribution, defined by $j_{Higgs}^{NL}(\Omega)=j^{NL}_{Higgs+BCS}(\Omega)-j^{NL}_{BCS}(\Omega)$, and the one coming from Eq.\ \pref{eq:japprox} for $\omega=0$.}
      
\label{fig3}                                                   
\end{figure}

{\bf FH in $s$-wave superconductors.} Fig. \ref{fig1} shows the nonlinear FH current $j^{NL}(\Omega)$ at charge density $n=0.5$ (results for $n=0.875$ are shown in Fig. \ref{fig3}) for different levels of disorder, parametrized as $\gamma/(2\Delta)$. We checked that the optical absorption sets in at a frequency $\Omega=2\Delta$, in agreement with the expectation of the $T=0$ Mattis-Bardeen theory in an $s$-wave superconductor. One clearly sees a pronounced peak in $j^{NL}(\Omega)$ at $\Omega=2\Delta$, which is strongly enhanced when amplitude fluctuations of the order parameter are included in the computation. This result is in agreement with the time-domain solution of the density-matrix problem reported in Ref.\ ~\cite{katsumi_Phys.Rev.Lett.24}, and with the computation of the FH by including disorder within the Born approximation reported in Ref.\ ~\cite{tsuji_prb26}. The novelty of this manuscript is that the numerical solution built upon a perturbative expansion in the frequency domain allows us to identify the most relevant process responsible for such enhancement, and to provide us with a guideline to explain it analytically. In the more general case of a driving pulse with a finite spectral width, the process is represented diagrammatically in Fig. \ref{fig2}. The third-order nonlinear current at $\omega+\Omega$ (dashed wavy line) is a function of three perturbing fields $\bA$ (solid wavy lines) at $\Omega, -\Omega+\omega, \Omega$. Each of them drives particle-hole electronic excitations,  denoted by the solid lines in the loops of Fig.\ \ref{fig2}, through single-photon processes. In the presence of amplitude fluctuations of the SC order parameter, the quasiparticle excitations exchange a Higgs fluctuation (dashed straight line) before recombining with the probe field that induces the generated current, denoted via the dashed wavy line. The full process can be written, with a sum implicit over the external Matsubara frequencies $i\Omega_i$, as $j^{NL}(i\Omega_{123})=\chi^{(3)}(i\Omega_i)A(i\Omega_1)A(i\Omega_2)A(i\Omega_3)$ where:
\begin{equation}
\label{chi3}
   \chi^{(3)}(i\Omega_i) =\chi_{\Delta;AA}(i\Omega_1,i\Omega_2)H(i\Omega_{12})\chi_{A;\Delta A}(i\Omega_{12},i\Omega_{3})
\end{equation}
where $i\Omega_{ij\dots}\equiv i\Omega_i+i\Omega_j+\dots$, $H(i\Omega)$ is the propagator of Higgs fluctuations, and we denoted by $\chi_{A;BC}$ a susceptibility representing the response of the physical quantity $A$ induced by a perturbation driven by the combined action of the fields $BC$. In the Matsubara formalism, the frequencies of the fields acting as perturbations are taken as positively entering the loop, while the outgoing frequency is associated with the field on which the response is computed. This assignment is crucial to set causality relations once the analytical continuation from the positive Matsubara frequency to (positive or negative) real frequencies is performed, leading to  different retarded response functions ~\cite{eliashberg_JETPLett.70,rostami_AnnalsofPhysics21}. More specifically, $\chi_{\Delta;AA}$ represents the variation of the SC order parameter after electromagnetic irradiation. In contrast, $\chi_{A;\Delta A}$ will correspond to a modification of the electronic susceptibility in the presence of a modulation of the amplitude of the order parameter. To get an intuition about the behavior of the nonlinear kernel in Eq.\ \pref{chi3} we can make an analytical estimate based on the Lehmann representation of the response function on the quasiparticle basis $E_\alpha$. More specifically, we assume, in the spirit of Anderson's theorem~\cite{anderson_JournalofPhysicsandChemistryofSolids59}, that pairing occurs between time-reversed states and spatial fluctuations of the order parameter can be neglected. As a consequence, $\sum_i \Delta_i c^\dagger_{i\down}c^\dagger_{i\up}\simeq \Delta\sum_\alpha c^\dagger_{\alpha\down}c^\dagger_{-\alpha\up}$, where $c_{\pm\alpha\sigma}$ annihilates the exact disorder eigenstates $\pm\alpha$ of the metal with energy $\xi_\alpha$ and spin $\sigma$, and the spectrum of Bogoliubov quasiparticles is given by  $\pm E_\alpha=\pm\sqrt{\xi_\alpha^2+\Delta^2}$.
Under these conditions, the most singular term in the left bubble $\chi_{\Delta;AA}(i\Omega_1,i\Omega_2)$ of Fig.\ \ref{fig2} can be simply written in the low-temperature limit (see ~\cite{suppl}) as:
\begin{align}
    \chi_{\Delta;AA}(i\Omega_1,i\Omega_2)=\sum_{\a\b}\frac{|V_{\a\b}|^2 \Delta_{\a\a
    }}{i\Omega_{12}}&\left[ \frac{f(E_\a)-f(-E_\b)}{E_{\a\b}+i\Omega_1}\right.\nn\\
\label{chileft}
    &+\left.\frac{f(E_\a)-f(-E_\b)}{E_{\b\a}+i\Omega_2}\right],
\end{align}
where $f(x)$ is the Fermi function, $V_{\a\b}$ represents the product of the matrix element of the current operator between the states $E_\alpha$ and $E_\beta$ times the Bogoliubov coherence factors, while $\Delta_{\alpha\alpha}\equiv \Delta/E_\alpha$ is the matrix element of the pairing operator $\hat\Delta$, and $E_{\a\b}=E_\a+E_\b$. In Eq.\ \eqref{chileft} we included only the most singular interband terms relevant as $T\rightarrow 0$, while additional non-singular contributions are detailed in ~\cite{suppl}. 
The right bubble $\chi_{A;\Delta A}(i\Omega_{12},i\Omega_{3})$ can be obtained from Eq.\ \pref{chileft} by replacing $i\Omega_1\ra i\Omega_3$ and $i\Omega_2\ra -i\Omega_{123}$, in agreement with the assignment explained above. Indeed, in this case one measures a current in response to a perturbation set by a photon field and a Higgs field. This frequency swapping in Matsubara space leads to a different outcome of the analytical continuation to the real frequencies in each bubble, which explains the results shown in Fig.\ \ref{fig1}. The process depicted in Fig.\ \ref{fig2} is obtained by taking $i\Omega_1,i\Omega_3\ra \Omega+i\eta$ and $i\Omega_2\ra \omega-\Omega+i\eta$, such that $i\Omega_{12}\ra \omega+i\eta$ and $i\Omega_{123}\ra \Omega+\omega+i\eta$, where $\eta>0$ is a vanishing positive quantity to ensure the analytical continuation of the retarded response function~\cite{eliashberg_JETPLett.70,gorkov_Sov.Phys.JETP69,
rostami_AnnalsofPhysics21}. When  $\omega\ra 0$ one obtains a nonlinear FH contribution. In this limit, the two response functions appearing in Eq.\ \pref{chi3} can be related to the {\em linear} current susceptibility, whose leading low-$T$ contribution can be  written as~\cite{suppl}:
\begin{equation}
\label{chi1}
    \chi_{jj}(i\Omega)=\sum_{\a\b}|V_{\a\b}|^2 \frac{f(E_\a)-f(-E_\b)}{E_\a+E_\b+i\Omega}.
\end{equation}
More specifically, neglecting irrelevant prefactors scaling as $\Delta_{\alpha\alpha}$ in Eq.\ \pref{chileft}
we obtain that:
\begin{align}
\chi_{\Delta;AA}(\omega;\Omega,\omega-\Omega)&\sim \frac{1}{\omega+i\eta } [\chi_{jj}(\Omega)-\chi_{jj}^*(\Omega)]\nn\\
&=\frac{2i{\rm Im}\chi_{jj}(\Omega)}{\omega+i\eta}, 
    \label{eqchid}
\end{align}
while the analytical continuation in the bubble $\chi_{A;\Delta A}(\omega+\Omega+i\eta; \omega+i\eta,\Omega+i\eta)\sim\chi_{\Delta;AA}(-i\Omega_{12}\to-\omega-i\eta;i\Omega_3\ra \Omega+i\eta, -i\Omega_{123}\ra -\Omega-\omega-i\eta)$ leads to:
\begin{align}
\chi_{A;\Delta A}(\omega+\Omega;\omega,\Omega)&\sim\frac{1}{\omega+i\eta} [\chi_{jj}(\Omega+\omega)-\chi_{jj}(\Omega)]\nn\\
&=\frac{\pd \chi_{jj}(\Omega)}{\pd \omega}. 
\label{eqchir}
\end{align}
A more accurate computation of Eq.\ \pref{eqchir}, taking into account the Bogoliubov transformation to the quasiparticle states in the definition \pref{chileft}, shows that $\chi_{A;\Delta A}$ should be expressed as the derivative of the linear susceptibility with respect to the order parameter (see ~\cite{suppl}). In both cases, qualitatively the same result is obtained, i.e.\ a response function which has singular behavior around the gap at $\Omega\sim2\Delta$ in an $s$-wave superconductor. By also including the diamagnetic contribution to the linear electronic susceptibility $\chi(\omega)$, one can  recast the nonlinear current in the form anticipated in Eqs.\ \pref{pnlsc}-\pref{eliash}, i.e.\
\begin{equation}
\label{eq:japprox}
  j^{NL}(\omega+\Omega)=\frac{\partial\chi(\Omega)}{\partial\Delta}\delta\Delta(\omega)A(\Omega), 
\end{equation}
 by the identification of the amplitude fluctuations in response to electromagnetic radiation with:
 \begin{eqnarray}
  \delta \Delta(\omega)&=&\chi_{\Delta;AA}(\Omega,\omega-\Omega)H(\omega)A(\Omega)A(\omega-\Omega)\nn\\
  \label{eq:delta}
  &\sim& \frac{2i{\rm Im}\chi(\Omega)}{\omega+i\eta}H(\omega)A(\Omega)A(\omega-\Omega),
 \end{eqnarray}
where the Higgs fluctuations are enhanced at $\omega=2\Delta$ in an $s$-wave superconductor~\cite{cea_Phys.Rev.B16,
tsuji_Phys.Rev.B15,schwarz_Phys.Rev.B20,tsuji_Phys.Rev.Research20,schwarz_NatCommun20,
silaev_Phys.Rev.B19,seibold_Phys.Rev.B21}. Eq.\ \pref{eq:delta} can be used to interpret the enhancement of the FH signal at $\Omega=2\Delta$ in our monochromatic simulations, where $\omega=0$ and we compute $j^{NL}(\Omega,-\Omega,\Omega)$. In the $\omega\ra 0$ limit the THz field drives large $\delta \Delta(\omega=0)$ fluctuations of the order parameter due to the enlarged susceptibility $\chi_{\Delta;AA}$ in Eq.\ \pref{eqchid}, whose divergence is only cut-off if $\eta$ is taken as a finite number. We remark here that Eq.\ \pref{eq:delta} accounts only for the most diverging term contributing to $\chi_{\Delta;AA}$, which is activated at $\Omega>2\Delta$ and is related to absorption above the gap edge. The full expression in the SM ~\cite{suppl} also contains a finite, non-diverging contribution which originates instead from the \textit{real} part of $\chi(\Omega)$, and it is thus operative also in the subgap pumping regime.

In Fig.\ \ref{fig3} we compare the full numerical computation of the Higgs-only contribution, $j_{Higgs}^{NL}(\Omega)=j^{NL}_{Higgs+BCS}(\Omega)-j^{NL}_{BCS}(\Omega)$ with the contribution due only to the diagram shown in Fig.\ \pref{fig2}. More specifically, we extract from the numerical results themselves both 
the static fluctuations of the order parameter $\delta\Delta(\omega=0)$, shown in Fig.\ \ref{fig3}a, and 
$\partial\chi(\Omega)/\partial\Delta$, shown in Fig.\ \ref{fig3}b, and we use them to evaluate Eq.\ \pref{eq:japprox} for the parameter values of Fig.\ \ref{fig1}b. As shown in panel \ref{fig3}d the expression provided by Eq.\ \pref{eq:japprox} is in excellent agreement with the full result, showing not only the predominance of this process over the others but also the validity of the analytical estimate. In addition, the numerical results are also in good agreement with a full analytical computation of the product $\delta \Delta(0) (\pd \chi/\pd \Delta)$ based on the Zimmermann ~\cite{zimmermann_PhysicaC:Superconductivity91} approximation (see also ~\cite{suppl} for additional details), as shown in Fig.\ \ref{fig3}c. {For this check, we have replaced the summation over disorder eigenstates appearing e.g.\ in Eq.\ \pref{chileft} with an integral over energies weighted by a Lorentzian kernel, as originally done for the conductivity to keep a finite mean free path ~\cite{zimmermann_PhysicaC:Superconductivity91}.} Notice that $\delta \Delta(\omega)$ in Eq.\ \pref{eq:delta} is closely related to the so-called Eliashberg mechanism, which predicts a  superconductivity enhancement via THz radiation only in a narrow range of temperatures near $T_c$~\cite{eliashberg_JETPLett.70,derendorf_Phys.Rev.B24}. In our $T=0$ case $\delta \Delta(0)$ in Fig.\ \ref{fig3}a is
{\em negative}, i.e. the static order parameter is suppressed by the strong THz pulse. Nonetheless, such a suppression brings the system in a non-equilibrium configuration that can only relax back if finite energy scattering channels are allowed, as we will discuss below. As a final remark, we stress that the agreement between Eq.\ \pref{eq:japprox} and the exact numerical results improves with decreasing disorder, i.e.\ in the case $\gamma/(2\Delta)<1$.


{\bf Displacive driving of the Higgs mode}
The mechanism of nonlinear FH enhancement encoded in Eq.\ \pref{eq:japprox} involves static Higgs fluctuations: $H(0)$ in Eq.\ \pref{eq:delta} is a {\em finite} constant, and the enhancement of the nonlinear current is a consequence of the diverging susceptibility $\chi_{\Delta;AA}$ as $\omega\ra 0$, see Eq.\ \pref{eqchid}. A resonant contribution of the Higgs mode to the FH, analogous to the one discussed for the TH current~\cite{matsunaga_Science14,matsunaga_Phys.Rev.B17,kovalev_Phys.Rev.B21,chu_NatCommun20,chu_NatCommun23,yuan_Sci.Adv.24,grasset_npjQuantumMater.22,kim_Sci.Adv.24}, is in principle allowed by the combination $i\Omega_1,i\Omega_2\ra \Omega+i\eta$ and $i\Omega_3\ra -\Omega+i\eta$ in Eq.\ \pref{chi3}. In this case the amplitude fluctuations are driven at the sum-frequency combination of the incoming photons at $2\Omega$, i.e. $\chi^{(3)}(\Omega,\Omega,-\Omega)\propto H(2\Omega)$, that is diverging when $2\Omega=2\Delta$. However, one can show~\cite{suppl} that in this case the susceptibilities are finite, while the Higgs divergence at $2\Delta$ scales as $\sim 1/\sqrt{\eta}$, reminiscent of the singularity found in the clean limit~\cite{cea_Phys.Rev.B16,
tsuji_Phys.Rev.B15,schwarz_Phys.Rev.B20,tsuji_Phys.Rev.Research20,schwarz_NatCommun20,
silaev_Phys.Rev.B19,seibold_Phys.Rev.B21}. Overall, these results explain why in the limit $\eta\ra0$ the resonant excitation of the Higgs mode is subleading with respect to the $1/\eta$ divergence of Eq.\ \pref{eq:delta}, accounting for the negligible FH enhancement at $2\Omega=2\Delta$ seen in the numerical simulations of Fig.\ \ref{fig1}a. 

The mechanism described here for the FH enhancement is analogous to the one discussed in the past as the crossover from impulsive to displacive stimulated Raman excitation of coherent phonons in semiconductors~\cite{merlin_ssc97,merlin_prb02}. This becomes operative when the energy of the visible light pulse exceeds the band gap energy leading to absorption. Indeed, at $T=0$ the $1/\eta$ divergence in $\chi_{\Delta;AA}\neq 0$ occurs only at $\Omega>2\Delta$, where {\em real} interband transitions between the SC bands are possible, making ${\rm Im}\chi(\Omega)\neq 0$ and allowing for a large $\delta \Delta(\omega)$ in Eq.\ \pref{eq:delta}. In the Lehmann representation of the susceptibilities entering the process of Fig.\ \ref{fig2} one can assign~\cite{suppl} the enhancement of the nonlinear response to the displacive driving of a collective mode (amplitude fluctuations in a superconductor, lattice vibrations in a semiconductor) able to sustain the pump-induced excited state without decaying in energy (at $\eta=0$). This condition can be understood by computing the power absorbed in the system under the same current perturbation $H_{int}=-\hat j\cdot \bA(t)$. It is a known result~\cite{fabrizio_book} that the power scales with the time derivative of the imaginary part of the $\chi_{jj}$ susceptibility, which in turn is only operative when real interband transitions are allowed. This leads to a time-dependent average energy $U(t)$, whose Fourier transform can be expressed~\cite{suppl} via a three-point correlation function $\chi_{H;AA}$ having the same structure of Eq.\ \pref{chileft}, provided that $\Delta_{\alpha\alpha}\ra E_\alpha$. As a consequence, $U(\omega)$ as $\omega\ra 0$ has the same structure as Eq.\ \pref{eq:delta}, i.e.\
\begin{equation}
\label{eq:cons}
    U(\omega)=\frac{2i\text{Im} \chi_{jj}(\Omega)\Omega}{\omega+i\eta}A(\Omega)A(\omega-\Omega).
\end{equation}
In the limit $\eta=0$ and $\Omega>2\Delta$, $U(\omega\ra 0)$ diverges, signaling an unbounded increase of energy due to optical absorption. On the other hand, if we apply this result to the energy $U_{BCS}$ computed on the Bogoliubov eigenstates allowing for a finite $\eta$ due to inelastic effects, this value will cut off the long-time absorption, allowing one to identify this quantity as the energy-relaxation rate of the quasiparticles. As further discussed in ~\cite{suppl}, the $1/\omega$ behavior of Eq.\ \pref{eq:cons} follows from the $1/\Omega_{12}$ of Eq.\ \pref{chileft}. This is in turn a general consequence of the fact that excited quasiparticles recombine in a collective mode associated with a {\em conserved} electronic quantity in $H_{BCS}$, i.e. whose operator is diagonal in the quasiparticle basis.

\begin{figure}
  \includegraphics[width=\columnwidth,clip=true]{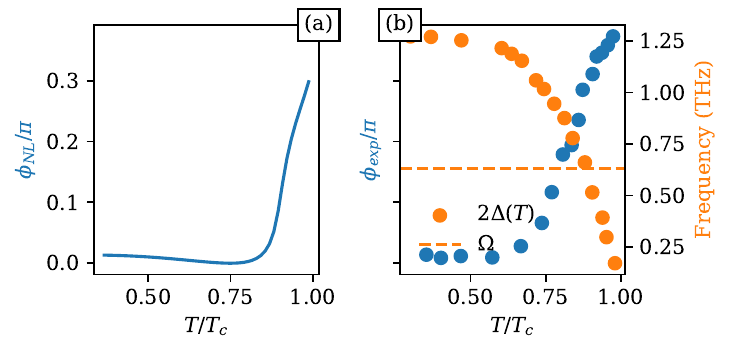}
  \caption{(a): Theoretical phase shift on the nonlinear FH simulated in the Zimmermann approximation (see SI) using $\gamma/(2\Delta)=1.5$ and $\eta/(2\Delta)=0.05$. (b): Experimental phase shift (blue) of the nonlinear first harmonic measured on a NbN sample with $T_c\approx14.9$ K and $2\Delta(T=0)\approx1.2$ THz through a narrowband THz pulse with central frequency $\Omega\approx0.63$ THz (orange dashed line). The orange dots represent twice the SC gap $2\Delta(T)$ {against temperature} measured via {THz} linear spectroscopy in Ref.\ ~\cite{katsumi_Phys.Rev.Lett.24}. }     
\label{fig4}
\end{figure}  

{\bf Fingerprints of the displacive Higgs excitation in time-resolved protocols.} A first experimental test of the mechanism encoded in Eq.\ \pref{eq:japprox} is the temperature evolution of the FH signal induced by a multicycle (almost monochromatic) light pulse at $\Omega$. The enhancement of the FH intensity at the optical gap can be observed by tuning $\Delta$ with temperature such that $\Omega=2\Delta(T^*)$, as already noticed in Ref.\ ~\cite{katsumi_Phys.Rev.Lett.24}. Here we prove that this effect is accompanied by a shift around $T^*$ of the nonlinear FH phase $\phi_{NL}=\text{arg}(j^{NL}(\Omega))$. Experimentally, one can extract the relative phase $\phi_{exp}$ between the transmitted field in linear response and the nonlinear FH signal~\cite{katsumi_Phys.Rev.Lett.24,suppl}. Since in the thin-film limit they both scale with the corresponding current, and $j^{NL}(\Omega)\sim |A(\Omega)|^2A(\Omega)$ has the same phase of the linear response, $\phi_{exp}$ is a good proxy to the FH phase $\phi_{NL}$. In Fig.\ \ref{fig4}b we report the measured value on NbN of $\phi_{exp}$ and of the SC gap $\Delta(T)$. According to Eq.\ \pref{eq:japprox}, since $\delta\Delta(\omega=0)$ is a real quantity, the complex phase $\phi_{NL}$ directly tracks the phase of $\pd \chi/\pd \Delta$, simulated in Fig.\ \ref{fig4}a for the experimental conditions of Ref.\ ~\cite{katsumi_Phys.Rev.Lett.24}. In general, the phase approaches zero when $\Omega\ll 2\Delta$, due to the lack of absorption in the linear response, and it approaches a value at most of order $\pi/2$ when $\Omega\gg 2\Delta$, depending in general on the disorder level, see Fig.\ \ref{fig4}b. 

A second experimental signature of the order-parameter modulation at zero frequency occurs in time-resolved protocols. To keep the analogy with the well known case of the displacive excitation of coherent phonons ~\cite{merlin_ssc97,dresselhasu_prb92,merlin_prb02}, we will consider the typical scheme of one-dimensional pump-probe measurements~\cite{matsunaga_Phys.Rev.Lett.13,matsunaga_Science14}. In this case the observable is the variation of the transmitted electric field $\delta E^{out}(\tau)$ through a thin SC film in the presence of a pump, as a function of the pump-probe delay $\tau$. For thin films $\delta E^{out}(\tau)\propto j^{PP}(\tau)\equiv j^{NL}(t=t_0,\tau)$ where the non-linear current is computed  at a fixed acquisition time $t=t_0$. The pump-probe scheme accesses a non-linear response that is linear in the detection field $A_0(t)$ and quadratic in the stronger pump pulse $A_\tau(t+\tau)$, whose envelope is centered at $t=-\tau$ with respect to the probe. The step-by-step correspondence between nonlinear processes represented in a diagrammatic language and the measured experimental signal as a function of $(t,\tau)$ is detailed in Ref.\ ~\cite{suppl}, along the lines  recently provided in Ref.\ ~\cite{fiore_prb26} within the context of 2DCS. We will consider for simplicity the case where the pump pulse is narrowband at $\Omega$, modeled as $A_\tau(t)=e^{-\bar{\sigma}^2t^2/2}\cos\Omega t$, $\bar{\sigma}=\sigma/(2\pi)$, and the probe one is simplified as extremely broadband. Since the detection process has to do with the right box of the diagram in Fig.\ \ref{fig2}, one can intuitively understand that in this situation the  $\chi_{A;\Delta A}$ function gives a smooth contribution, as shown explicitly in Ref.\ ~\cite{suppl}. On the other hand the pump process, connected to the left part of the diagram, keeps trace of the singular behavior of the $\chi_{\Delta;AA}$ in the frequency $\omega_\tau$ corresponding to the Fourier transform of the time delay. More specifically, one can show~\cite{suppl} that  $j^{PP}(\omega_\tau)$ reads:
\begin{align}
\label{eq:s}
    j^{PP}(\omega_\tau)&\simeq H(\omega_\tau)\nonumber\\
    &\times\int d\omega\,\chi_{\Delta;AA}(\omega,\omega_\tau-\omega)A_\tau(\omega)A_\tau(\omega_\tau-\omega),
\end{align}

To get intuition on the outcome of Eq.\ \pref{eq:s}, one should consider that whenever the $\omega$ dependence of $\chi_{\Delta;AA}$ is smooth over the spectral range of the pump pulse the integration over $\omega$ reconstructs $A_\tau^2(\omega_\tau)=\int d\omega A_\tau(\omega)A_\tau(\omega_\tau-\omega)$, i.e.\ the Fourier transform of $A_\tau^2(t)$. In other words, one combines two photons of the pump pulse to drive the amplitude mode at the frequency $\omega_\tau$, with a strength depending on the overall scaling $\chi_{\Delta;AA}\sim 2i\text{Im}\chi_{jj}(\Omega)/(\omega_\tau+i\eta)$ of the $\chi_{\Delta;AA}$ susceptibility, see Eq.\ \pref{eqchid} above, and on the square-root divergence of the Higgs propagator at $2\Delta$,  $H(\omega_\tau)\sim 1/\sqrt{2\Delta-\omega_\tau}$. Depending on the central frequency $\Omega$ and on the spectral width $\sigma$ of the pump one can have only one contribution or both, as demonstrated in Fig.\ \ref{figpp} where we show a numerical simulation of  $j^{PP}(\tau)$ obtained via an analytical approximation to the response function.  See ~\cite{suppl} for further details. In panel (a) we show the spectrum of the $A_\tau(\omega)$ (solid lines) along with  the one of $A_\tau^2(\omega)$ (dashed lines). For a pump centered at $\Omega$,  $A_\tau^2(\omega)$ has its main spectral components at $\omega\simeq 0$, corresponding to a  difference-frequency process with two photons of opposite frequency, and at $\omega\simeq 2\Omega$, corresponding to a sum-frequency process with two photons near the same frequency. When $\Omega<2\Delta$ (blue curves) the displacive mechanism is absent. In this case, the Higgs mode can only be driven by the optically forbidden sum-frequency process $2\Omega\simeq 2\Delta$.  The current as a function of $\tau$ {(at fixed $t$) shows $2\Delta$ oscillations about zero (see Fig.\ \ref{figpp}b), with a $1/\sqrt{\tau}$ decay in time characteristic of the amplitude mode~\cite{volkov_Sov.J.Exp.Theor.Phys.74,yuzbashyan_prl06}. 
{ Note that in principle $2\Delta$ oscillations could also arise from the  purely quasiparticle contribution~\cite{cea_Phys.Rev.B16}, not discussed here, that is dominant in the absence of disorder. In the clean system a sizeable photo-induced dynamics of the order parameter $|\delta\Delta(\tau)|$ emerges only when the $\chi_{\Delta;AA}$ response is evaluated at finite momentum $\mathbf{q}$ of the incoming radiation~\cite{papenkort_prb07,papenkort_prb08,krull_prb14}.   In disordered systems, one can keep the pump at $q=0$ and an effective momentum is instead provided by scattering with point-like impurities, making ultimately the Higgs contribution at $2\Delta$ larger than the quasiparticle one at strong disorder~\cite{silaev_Phys.Rev.B19,tsuji_Phys.Rev.Research20,seibold_Phys.Rev.B21,benfatto_Phys.Rev.B23,haenel_Phys.Rev.B21}, which is the case addressed here. }

\begin{figure}
  \includegraphics[width=\columnwidth,clip=true]{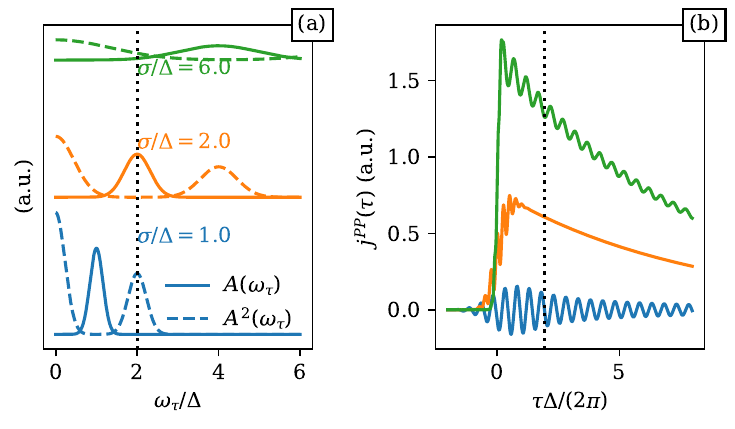}
  \caption{(a): The spectral envelope $A(\omega_\tau)$ of the pump pulse (solid lines), centered at $\Omega$, and the squared pulse $A^2(\omega_\tau)$ (dashed lines), centered at the sum-frequency combination $2\Omega$ and at the difference-frequency combination around zero. The Higgs mode can be excited at $2\Delta$ either via a sum-frequency process (blue  lines), or via a difference-frequency process (green lines), as in the diagram described in Fig.\ \ref{fig2}.  The displacive character depends on the value of $\Omega$ with respect to the onset of quasiparticle excitations (black dotted line), making it ``impulsive'' (blue lines) or ``displacive'' (orange and green lines). (b): Time evolution of the pump-probe signal for varying pump frequency $\Omega$. When $\Omega<2\Delta$ (blue line) only $2\Delta$ oscillations around zero are observed (impulsive limit). As $\Omega>2\Delta$ (orange and green lines) a displacive component emerges, with a finite time-integrated value, and $2\Delta$ oscillations have a relative $\pi/2$ shift with respect to the impulsive case (black dotted line is a guide to the eye). In the orange case the spectral weight of $A^2(\omega_\tau)$ at $2\Delta$ is vanishing (see panel (a)), and Higgs oscillations disappear. Curves are rescaled to an overall prefactor to be shown on the same panel. }     
\label{figpp}
\end{figure}

When $\Omega=2\Delta$ (orange line) displacive excitation becomes operative, as evidenced by the large increase of the signal followed by a slow decay. However, in this case $A_\tau^2(\omega)$ only has appreciable intensity at 0 and $4\Delta$ (see dashed orange line in Fig.\ \ref{figpp}a), so $2\Delta$ Higgs oscillations cannot be triggered. The only $2\Omega$ oscillations come from the pump, which lasts only for the time window of the pump itself. In this regime, the one-dimensional spectrum as a function of $\omega_\tau$ of Eq.\ \pref{eq:s} scales as:
 \begin{equation}
\label{eq:jpp}
    j^{PP}(\omega_\tau\approx0)\sim \frac{iH(0)}{\omega_\tau+i\eta}A_\tau^2(\omega_\tau)
\end{equation}
whose Fourier transform reads approximately:
\begin{equation}
\label{eq:jppt}
    j^{PP}(\tau)\sim H(0)e^{-\eta \tau}\int_{-\infty}^\tau dt A_\tau ^2(t)
\end{equation}
where we also assumed that the duration of the pump pulse $A_\tau(t)$ is significantly shorter than $1/\eta$, as is the case in Fig.\ \ref{figpp} where $\eta/\Delta=0.01$. 
 Eq.\ \pref{eq:jppt} establishes a direct correspondence between the mechanism of rectified  displacive  excitations of the Higgs mode, that is cut-off in energy by the inelastic scattering rate $\eta$, and the emergence of a long-lasting signal in pump-probe protocols, that is cut-off in time by $1/\eta$.  Being a non-oscillating signal, this appears in the 2DCS spectra along the $\omega_\tau = 0$ line and is usually called a pump-probe (PP) response in the 2DCS literature~\cite{mahan_00,fiore_prb26}.  This scheme has been indeed recently used to track the evolution of the energy relaxation rate $\eta$ in a variety of SC and non-SC systems~\cite{barbalas_Phys.Rev.Lett.25,chaudhuri_25}.

{
Finally, for a broader pump with $\Omega > 2\Delta$, the difference-frequency process carries sufficient spectral weight at $2\Delta$ (dashed green line in Fig.\ \ref{figpp}a) to displacively drive both the rectified and the $2\Delta$ oscillatory components of the Higgs mode (solid green line in Fig.\ \ref{figpp}b). Notably, we observe a $\pi/2$ phase shift of these $2\Delta$ oscillations compared to the impulsive case. This situation is indeed the closest to the conventional displacive excitation of coherent phonons in opaque materials via visible femtosecond pulses~\cite{dresselhaus_apl90,merlin_prb02,bossini_prb25}. In that case, a pump pulse with a difference-frequency peak centered at $\omega = 0$ has a broad THz-range bandwidth, driving large phonon oscillations with an analogous $\pi/2$ phase shift~\cite{merlin_prb02}. In non-linear experiments using broadband THz pump pulses, the zero-frequency displacive effect typically coexists with $2\Delta$ oscillations, as illustrated by the green curve in Fig.\ \ref{figpp}b, which agrees with the experimental observations in NbN~\cite{matsunaga_Phys.Rev.Lett.13}. Conversely, in narrowband experiments ~\cite{matsunaga_Science14,giorgianni_Nat.Phys.19,yang-natphot19,wangNL_prl25,kim_Sci.Adv.24} the oscillations occur at $2\Omega$ (orange line in Fig.\ \ref{figpp}b), and the displacive behavior only emerges when the pump central frequency exceeds the threshold ($\Omega > 2\Delta$). While previous literature on both narrow- and broadband experiments has generically attributed this displacive signal to condensate depletion ~\cite{demsar_prl13,demsar_prl23,matsunaga_Phys.Rev.Lett.13,matsunaga_Science14}, our work provides and quantifies a specific microscopic mechanism. Indeed, our numerical computation of the FH current reveals that although full paramagnetic processes involving only quasiparticle excitations scale as $1/\eta$~\cite{suppl}, their contribution in $s$-wave superconductors is subdominant to the Higgs-mediated process across the explored disorder range (see Fig.\ \ref{fig1}). This dominance allows us to unambiguously identify the displacive non-linear signal as a signature of static order-parameter modulation. The same non-linear process also underlies the spectral structures around $2\Delta$ recently observed in 2DCS maps~\cite{katsumi_Phys.Rev.Lett.24}.  A more detailed analysis of this process, utilizing the theoretical framework of Ref.\ ~\cite{fiore_prb26}, is reserved for future work ~\cite{fiore_future}.
}

\begin{figure}
  \includegraphics[width=\columnwidth, clip=true]{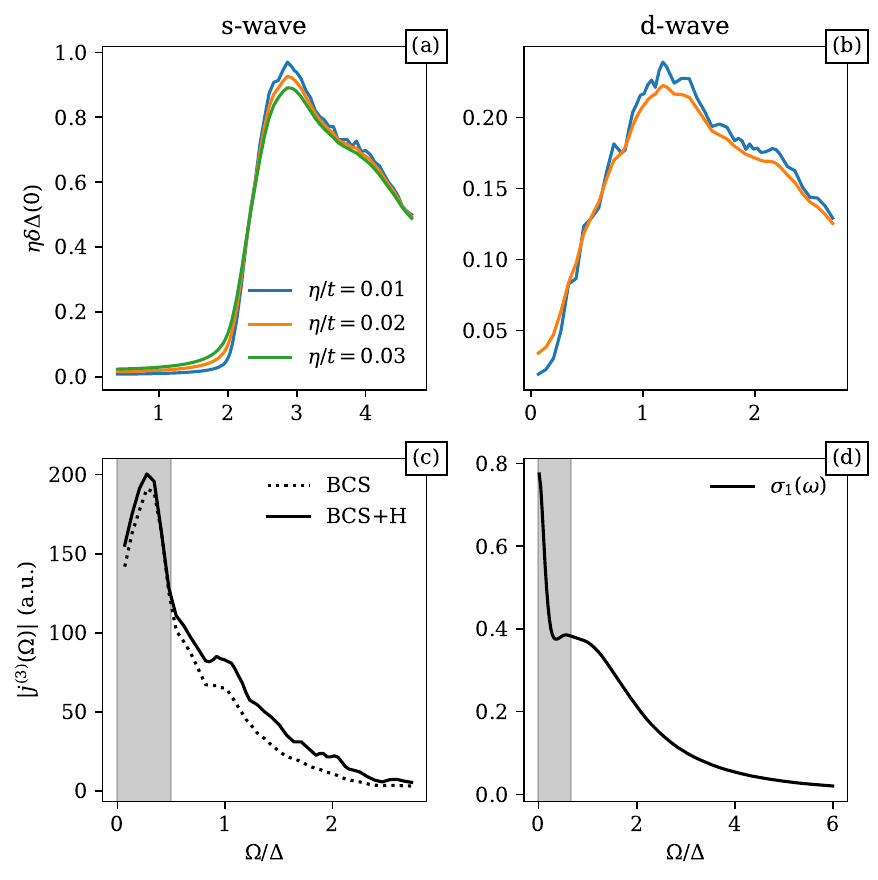}
  \caption{Scaling analysis of the pump-induced variation of the order parameter $\delta\Delta(0)$ for different values of $\eta$ as a function of pump frequency for (a) $s$-wave and (b) $d$-wave superconductors. In both cases the results are compatible with $\sim1/\eta$ scaling. (c) Amplitude of the nonlinear FH at BCS level (dotted line) and with the inclusion of Higgs fluctuations  (solid line)  for $d$-wave superconductors. Parameters here: $n=0.875$, $J/t=1$, $t'/t=-0.2$, impurity concentration $c=1$. The shaded region signals where finite-size effects affect the results. (d) Linear optical conductivity in the $d$-wave case. The smooth behavior of the $\chi(\Omega)$ at twice the gap maximum explains the lack of a sharp Higgs contribution to the FH current, shown in panel (c).}
\label{fig5}                                                   
\end{figure}  

{\bf Displacive Higgs behavior and FH for $d$-wave superconductors} The numerical approach based on the model \pref{eq:ham} can be extended to the case of a SC order parameter with $d$-wave symmetry. We include a next-nearest-neighbors hopping $t'$ and, instead of the attractive on-site interaction $U$, we add an interaction term $H_{int}=J\sum_{\langle ij\rangle}\left\lbrack \hat{\bf S}_i \hat{\bf S}_j -\frac{1}{4}\hat{n}_i \hat{n}_j\right\rbrack$, that can be decoupled at mean field in the $d$-wave pairing channel. We set $J/t=1$ and $t'/t=-0.2$ to mimic the case of cuprate superconductors. The main consequence of the presence of nodes in the order parameter is the persistence of low-energy excitations, that make the linear absorption smooth in frequency, see Fig.\ \ref{fig5}d. This removes both the sharp onset of $\delta \Delta(0)$ (Fig.\ \ref{fig5}b)  and the divergence of $\pd \chi/\pd \Delta$ at $\Omega=2\Delta$. As a result, the nonlinear FH current, shown in Fig. \ref{fig5}c, displays a smooth increase in frequency, and the inclusion of amplitude fluctuations barely modifies the response. On the other hand, a displacive effect on the Higgs is still operative in a broad energy range below $2\Delta$. In Fig.\ \ref{fig5}b we compare the simulated $\delta\Delta(0)$ as a function of the driving pulse frequency $\Omega$ at various $\eta$ values, and we show that the data scale as $1/\eta$, as expected on the basis of Eq.\ \pref{eliash}. Such a scaling holds both for the $s$-wave and the $d$-wave case, as shown in panel (a) and (b), respectively. We notice that a recent work ~\cite{dzero_cm26} analysed the $\delta\Delta(\omega)$ response in clean $d$-wave superconductors using the Eilenberger equation, finding no enhancement of the order parameter close to the critical temperature. However, we stress that our numerical calculations are performed including disorder effects at $T=0$, providing thus a complementary information with respect to that work. 

{\bf Conclusions} The present manuscript provides a unified microscopic view on a set of nonlinear phenomena that are usually ascribed to different processes, and whose theoretical explanation resides in the ability of the nonlinear optical response to access the parametric dependence of the linear optical susceptibility on a collective order parameter. Whenever the order is associated with an operator that contains a finite diagonal projection on the basis of the energy eigenstates of the system, the fluctuation-dissipation theorem leads to a novel paradigm: the absorption (dissipation) is converted in a static distortion of the collective order, rather than its fluctuations. Such a displacive effect, made possible by optical quasiparticle excitations above the bandgap, potentially persists over extremely long time scales, since the system can recover the equilibrium condition only via energy-relaxation processes. 
{This displacive paradigm can be recast in the language of the ponderomotive potential recently developed for driven collective coordinates \cite{sun_prb24, sun_prb26}: the static shift of the gap is a ponderomotive displacement of the order parameter, complementary to the dirty-superconductor calculation of Ref.\ \cite{sun_prb26}. Here we access this shift through its imprint on the nonlinear optical response, both in the first-harmonic response and in the long-lived time-resolved signal.}


As we have shown, in superconductors such a mechanism is always operative via a static deformation of the order-parameter amplitude, for both $s$-wave and $d$-wave symmetry of the order parameter. In the $s$-wave case we identified several experimental signatures of this effect. The first one is the enhancement of the intensity of the nonlinear FH current $j^{NL}(\Omega)$ at $\Omega=2\Delta$, observed in superconducting NbN~\cite{katsumi_Phys.Rev.Lett.24}, that we show here to occur along with an approximately $\pi/2$ shift of its phase $\phi_{NL}=\text{arg}(j^{NL}(\Omega))$, due to the emergence of an imaginary part of $\pd \chi/\pd \Delta$ in Eq.\ \pref{pnlsc} across the optical gap. This prediction is confirmed by experimental estimates of $\phi_{NL}$ obtained by measuring the relative phase between the FH signal and the linear one. In pump-probe protocols the displacive behavior manifests as a signal with non-zero average value over the delay time. This effect,  which has been usually ascribed to a consequence of quasiparticle absorption ~\cite{demsar_prl13,demsar_prl23,matsunaga_Phys.Rev.Lett.13,matsunaga_Science14}, is here identified and quantified in terms of a specific mechanism related to a static modulation of the order-parameter amplitude.   
In the $d$-wave case, the lack of a sharp absorption edge at twice the gap maximum removes the strong Higgs contribution in the FH at $\Omega=2\Delta$. As a consequence, we do not expect experimentally a strong enhancement of the FH signal in temperature at the crossover $\Omega=2\Delta(T^*)$. This result is analogous to the  previous finding for the TH intensity ~\cite{benfatto_Phys.Rev.B23}, that has been shown to increase monotonically below $T_c$ without any clear resonance~\cite{chu_NatCommun20,chu_NatCommun23,kim_Sci.Adv.24,yuan_Sci.Adv.24}. On the other hand, the Higgs mode can still provide a natural channel for a long-lasting pump-probe signal, cut-off by energy dissipation processes at $1/\eta$. Interestingly, in a recent report~\cite{chaudhuri_25} such a long-lasting signal has been  systematically investigated in $d$-wave cuprates at several doping levels. The displacive-light decay has also been observed to persist above the critical temperature, with a smooth evolution throughout $T_c$. In light of our findings, one can argue about the existence in these complex systems of a  collective electronic order able to sustain the optical absorption already above $T_c$, making 2DCS a preferential tool to explore unconventional relaxation mechanisms in correlated electron systems. 


{\bf Acknowledgements} L.B. and I.Z. acknowledge financial support by European Union under the project MORE-TEM ERC-SYN (Grant No. 951215); by Sapienza University under the projects Ateneo (Grants No. RM123188E357C540, No. RP124190A63FAA97, and No. RM125199BD4DE0C4); and by the Italian Ministry of Education, University, and Research under Project No. PRIN2022-CoInEx (2022WS9MS4). G.S. acknowledges support from the
Deutsche Forschungsgemeinschaft under SE 806/20-1. K.K. acknowledges funding from the U.S. National Science Foundation (DMR-2442520).  The work at JHU was supported by the National Science Foundation (DMR 2226666) and the Gordon and Betty Moore Foundation’s EPiQS Initiative through Grant No. GBMF9454.

\bibliography{2d_bib.bib,phd_thesis_merged.bib}

\end{document}